\documentclass[10pt,aps,prd,twocolumn,nofootinbib,superscriptaddress,preprintnumbers]{revtex4-2}

\usepackage{amsmath,amssymb,bm}
\usepackage[dvipdfmx]{graphicx}
\graphicspath{{./}}
\usepackage{float}
\usepackage{color}
\usepackage{hyperref}
\usepackage{physics}

\newcommand{\rd}{{\rm d}}
\newcommand{\cO}{{\cal O}}
\newcommand{\cF}{{\cal F}}
\newcommand{\cG}{{\cal G}}

\newcommand{\cK}{{\cal K}}
\newcommand{\GB}{R_{\rm GB}^2}
\newcommand{\be}{\begin{equation}}
\newcommand{\ee}{\end{equation}}
\newcommand{\ba}{\begin{eqnarray}}
\newcommand{\ea}{\end{eqnarray}}

\begin{document}

\preprint{YITP-26-86, WUCG-26-06}

\title{Static regular black holes in Horndeski theories: \\
analytic no-go and nonanalytic obstructions}

\author{Antonio De Felice}
\affiliation{Center for Gravitational Physics and Quantum Information, 
Yukawa Institute for Theoretical Physics, Kyoto University, Kyoto 606-8502, Japan}

\author{Shinji Tsujikawa}
\affiliation{Department of Physics, Waseda University, Shinjuku, Tokyo 169-8555, Japan}

\date{\today}

\begin{abstract}

Regular black holes in Horndeski theories must have stable horizons and
regular centers. We study static, spherically symmetric, asymptotically flat
configurations with a time-independent scalar. The horizon branch on which
the scalar kinetic term $X$ remains nonzero is generically obstructed by
divergent propagation speeds or ghost/gradient instabilities, aside from
special degeneracies. On the regular branch, where $X$ vanishes at the
horizon, analyticity at the relevant $X=0$ endpoints reduces the leading
scalar equation to finite sets of Taylor coefficients. For nondegenerate
shift-symmetric theories this gives a nonperturbative current no-hair
theorem: the scalar is constant and the metric is Schwarzschild, hence
centrally singular for nonzero ADM mass. For non-shift-symmetric
positive-power couplings, the corresponding exclusion applies to the
perturbative branch continuously connected to Schwarzschild. We also classify
marginal nonanalytic departures: covariant regularity fixes the
scalar--Gauss--Bonnet chain as the unique marginal nonanalytic completion.
Hairy black holes in this completion evade the analytic current step but
remain centrally singular.

\end{abstract}

\maketitle

\section{Introduction}

Regular black holes (BHs) sharply test whether spacetime singularities can be
avoided.  In general relativity (GR), the Schwarzschild solution has a central
singularity, and Penrose's theorem shows that singularities arise under
standard assumptions \cite{Penrose:1964wq}.  Early constructions such as the
Bardeen metric, de Sitter-core models, quantum-gravity-inspired geometries,
and black-bounce spacetimes provide explicit regular centers
\cite{Bardeen:1968,Dymnikova:1992ux,Hayward:2005gi,Bonanno:2000ep,Nicolini:2005vd,Simpson:2018tsi,Maeda:2021jdc}.
These metric-level constructions, however, do not identify a classical theory
realizing dynamically viable regular BHs.   Whether such objects
arise from consistent classical dynamics remains open.

The vector--tensor sector provides a benchmark.  In Einstein gravity coupled to
nonlinear electrodynamics (NED), regular spherical metrics can be constructed
by choosing the NED Lagrangian
\cite{Ayon-Beato:1998hmi,Ayon-Beato:1999kuh,
Bronnikov:2000vy,Dymnikova:2004zc,Ansoldi:2008jw,Balart:2014cga,
Balart:2014jia,Fan:2016hvf,Rodrigues:2018bdc}.
Perturbation analyses of Refs.~\cite{DeFelice:2024seu,DeFelice:2024ops} showed
that nonsingular electric and magnetic NED BHs generically suffer an angular
Laplacian instability near the regular center.  Regular BHs also arise in pure
higher-curvature gravity with infinite towers
\cite{Bueno:2024dgm,Bueno:2024zsx,Bueno:2024eig}.  They contain no
fundamental four-dimensional Horndeski scalar; their spherical reductions only
yield effective two-dimensional Horndeski descriptions, distinct from the
scalar-tensor theories studied here.

We consider four-dimensional Horndeski theories \cite{Horndeski:1974wa} with a
static scalar profile, $\phi=\phi(r)$.  Time-dependent shift-symmetric branches,
$\phi=qt+\psi(r)$ \cite{Babichev:2013cya,Kobayashi:2014eva}, and regular BHs in
beyond-Horndeski/DHOST theories
\cite{Chagoya:2018lmv,Babichev:2020qpr,Baake:2021qja}, are outside our scope.
Previous Horndeski no-hair and stability studies mostly allowed centrally
singular asymptotically flat BHs, rather than imposing both horizon stability
and center regularity.  Hui and Nicolis \cite{Hui:2012qt} proved no hair for
static shift-symmetric BHs under their assumptions.  The linear
scalar--Gauss--Bonnet (sGB) coupling is a known escape, supporting hairy
asymptotically flat BHs \cite{Sotiriou:2013qea}, but they are centrally singular. 
For static profiles with nonzero $X$ 
at the horizon radius $r_s$, Refs.~\cite{Minamitsuji:2022vbi,Minamitsuji:2022mlv} found generic ghost or
Laplacian instabilities, or divergent sound speeds, excluding the
nonminimal-derivative-coupling BHs of
Refs.~\cite{Rinaldi:2012vy,Anabalon:2013oea,Minamitsuji:2013ura} as stable
candidates. They also proved no hair for analytic positive-power couplings on
the small-coupling branch connected to Schwarzschild, without imposing a regular
center.  It remained unclear whether the $X(r_s)=0$ analytic branch can reach a
regular center, or whether a controlled nonanalytic completion can do so.

Here we close these gaps in the static-scalar sector.  The branch with
$X \neq 0$ at $r=r_s$ is obstructed by leading stability conditions, up to
special simultaneous degeneracies.  On the regular branch $X(r_s)=0$, 
nondegenerate shift-symmetric analytic theories obey a nonperturbative
current no-hair theorem: the scalar is constant and the metric is
Schwarzschild, so nonzero ADM mass entails the Schwarzschild central
singularity.  Non-shift-symmetric positive-power couplings are excluded on the perturbative
branch connected to Schwarzschild. Covariant regularity fixes
the marginal nonanalytic completion uniquely to the sGB chain, whose hairy
BHs evade the current reduction but remain centrally singular.

\section{Horizon branches}
\label{sec:horizonbranch}

We consider the most general four-dimensional Horndeski action
\ba
\hspace{-0.5cm}
{\cal S} &=&
\int \rd^4x\sqrt{-g} \biggl[ G_2-G_3 \Box\phi 
+G_4 R \nonumber \\
\hspace{-0.5cm}
& &
+G_{4,X}\left\{ (\Box\phi)^2
-\phi_{\mu\nu}\phi^{\mu\nu} \right\}
+G_5 G_{\mu\nu}\phi^{\mu\nu} \nonumber \\
\hspace{-0.5cm}
& &
-\frac{G_{5,X}}{6}
\left\{ (\Box\phi)^3
-3\Box\phi\,\phi_{\mu\nu}\phi^{\mu\nu}
+2\phi_\mu{}^\nu\phi_\nu{}^\rho
\phi_\rho{}^\mu \right \} \biggr],
\ea
where $g$ is the determinant of the metric tensor $g_{\mu\nu}$, 
$R$ is the Ricci scalar, $G_{\mu \nu}$ is the Einstein tensor,  
$G_i$ ($i=2,3,4,5$) are functions of $\phi$ and
$X=-\nabla_{\mu} \phi \nabla^{\mu}\phi/2$ with 
$G_{i,X}=\partial G_{i}/\partial X$, and 
$\phi_{\mu\nu}=\nabla_\mu\nabla_\nu\phi$.  
We consider the static, spherically symmetric background line element and a time-independent scalar-field profile,
\begin{equation}
\rd s^2=-f(r)\rd t^2+h^{-1}(r)\rd r^2+r^2\rd\Omega^2,
\qquad
\phi=\phi(r).
\label{metric}
\end{equation}
We assume a nonextremal outer horizon at
$r=r_s$, with $f(r_s)=h(r_s)=0$ and $f,h>0$ 
outside it.  The goal of this
section is to identify the generic near-horizon obstruction and the regular
branch used in the no-go proof.

Near the horizon, let
\begin{align}
f&=f_1\Delta+\cO(\Delta^2),
\qquad
h=h_1\Delta+\cO(\Delta^2),
\nonumber\\
X&=X_s+X_1\Delta+\cO(\Delta^2),
\qquad
\Delta\equiv r-r_s,
\label{horExpfhX}
\end{align}
with $f_1,h_1>0$.  
Since $X=-h\phi'(r)^2/2$, 
a finite, nonzero $X_s=X(r_s)$ requires
\begin{equation}
\phi'(r)=\frac{\phi_1}{2\sqrt{\Delta}}+\cO(\Delta^{1/2}),
\qquad
X_s=-\frac{h_1\phi_1^2}{8}<0\,.
\label{phiExpXs}
\end{equation}
Thus a finite $\phi'(r_s)$ gives $X_s=0$, while $X_s\ne0$ comes with the
coordinate derivative singularity \eqref{phiExpXs}. 
We now recall why the latter branch is generically
incompatible with horizon stability.

For the odd-parity gravitational perturbation, the radial propagation speed squared is
$c_{r,{\rm odd}}^2=\cG/\cF$ \cite{Kobayashi:2012kh}, where
\ba
\hspace{-0.4cm}
\cF &\equiv&
2G_4+h\phi^{\prime 2}G_{5,\phi}
-h\phi^{\prime 2}
\left(\frac{h'}{2}\phi'+h\phi''\right)G_{5,X},
\label{FGdef} \nonumber \\
\hspace{-0.4cm}
\cG &\equiv&
2G_4+2h\phi^{\prime 2}G_{4,X}
-h\phi^{\prime 2}
\left(G_{5,\phi}+\frac{f'h\phi'}{2f}G_{5,X}\right).
\ea
The ghost and radial-gradient stability conditions include $\cF>0$ and
$\cG>0$.  On the $X_s\ne0$ branch, $\cF$ is finite but
\begin{equation}
\cG=\frac{X_s h_1\phi_1 G_{5,X}(\phi_s,X_s)}{2\sqrt{r-r_s}}
+\cO(1),
\label{Gleading}
\end{equation}
where $\phi_s=\phi(r_s)$.  
Unless $G_{5,X}(\phi_s,X_s)=0$, 
the radial speed diverges 
at the horizon, signaling a breakdown
of the linear perturbative description 
and raising concerns about the
well-posedness of the Cauchy problem.

Even after imposing $G_{5,X}(\phi_s,X_s)=0$, 
the even-parity scalar mode gives
a further obstruction.  
Its radial squared speed behaves as
\cite{Minamitsuji:2022vbi}
\begin{equation}
c_{r2,{\rm even}}^2=
\frac{2h_1X_s\kappa_r}{\zeta_r(r-r_s)}+
\cO(1),
\label{cr2evenNH}
\end{equation}
where $\zeta_r$ is finite for a regular nondegenerate horizon and
\ba
\kappa_r &\equiv&
X_s r_s^2\left(2X_sG_{3,XX}-G_{3,X}\right) \nonumber\\
&&+r_s^2\left(3G_{4,\phi}-4X_s^2G_{4,\phi XX}\right)
+2X_s^2G_{5,XX} .
\label{kappardef}
\ea
Thus $\kappa_r\ne0$ gives a divergent scalar radial speed.  If $\kappa_r=0$
is also imposed, the no-ghost and angular-gradient conditions for even-parity
modes require $\cK>0$ and $B_2>0$, with $\cK$ and $B_2$ defined in
Refs.~\cite{Kase:2021mix,Kase:2023kvq,Minamitsuji:2022vbi}.  
Near the horizon,
\begin{equation}
\cF\cK B_2=
-\frac{4h_1^2X_s^4r_s^4\kappa^2}
{\zeta^2(r-r_s)^2}
+\cO\left((r-r_s)^{-1}\right),
\label{FKB2neg}
\end{equation}
where $\zeta$ is finite and
\begin{align}
\kappa &\equiv
G_4G_{4,XX}+G_{4,X}^2
-G_{5,\phi X}\left(G_4-X_sG_{4,X}\right)
\nonumber\\
&\quad
-G_{5,\phi}\left(2G_{4,X}+X_sG_{4,XX}-G_{5,\phi}\right).
\label{kappadef}
\end{align}
All functions and their derivatives in \eqref{kappardef} and \eqref{kappadef}
are evaluated at $(\phi_s,X_s)$.  For $\kappa\ne0$, the product
$\cF\cK B_2$ is negative close to the horizon, so at least one of the required
stability conditions fails.

The near-horizon branch with finite nonzero $X_s$ 
is therefore ruled out by the leading near-horizon 
stability conditions unless all three leading coefficients
vanish simultaneously at the horizon,
\begin{equation}
G_{5,X}(\phi_s,X_s)=0,
\qquad
\kappa_r=0,
\qquad
\kappa=0\,.
\label{simuldeg}
\end{equation}
The conditions in Eq.~\eqref{simuldeg} are not, by themselves, a stability
criterion. They are only local algebraic degeneracies at the horizon. 
Enforced at $(\phi_s,X_s)$ alone, they cancel the leading 
singular terms but constrain only finitely many Taylor coefficients of $G_i$. Treating Eq.~\eqref{simuldeg} as
an identity in $X$ is a much stronger structural requirement. 
The homogeneous part
of the $G_3$ condition then admits the fractional-power branch
$G_3\supset c_3(\phi)|X|^{3/2}$. This includes, for example,
\begin{equation}
G_3=c_3(\phi)|X|^{3/2},
\qquad
G_4=\frac{M_{\rm Pl}^2}{2},
\qquad
G_5=0\,,
\label{fractionalG3Example}
\end{equation}
where $M_{\rm Pl}$ is the reduced Planck mass. 
We use $|X|$ because the regular-BH discussion 
includes both sides of the horizon,
where the sign of $X=-h\phi'^2/2$ need not be the same.
For this example,
$2XG_{3,XX}-G_{3,X}=0$ away from $X=0$ on each fixed-sign branch. 
This realization is nonanalytic at the asymptotically flat vacuum $X=0$.

Thus Eq.~\eqref{simuldeg} cancels only the leading singular terms on the
regular near-horizon branch \eqref{horExpfhX}, for which $X-X_s$ is analytic in
$\Delta=r-r_s$.  If the leading coefficients are tuned to vanish, the next
orders of the near-horizon quadratic action involve 
higher derivatives of the functions $G_i$ and subleading 
background coefficients; generically the
pathology reappears at the first nonzero order, whereas an all-order
cancellation would represent a highly degenerate branch.
There are also exceptional $X_s\ne0$ branches outside this analytic near-horizon
class. In regularized 4-dimensional Einstein--Gauss--Bonnet (4DEGB) 
gravity \cite{Glavan:2019inb,Fernandes:2020nbq,Hennigar:2020lsl,Lu:2020iav,Kobayashi:2020wqy}, 
for instance, $X-X_s\propto\Delta^{1/2}$ near the horizon.  Such branches are not classified
by the algebraic conditions \eqref{simuldeg} and require a separate
perturbative analysis.  In the known 4DEGB case, the separate perturbative analysis rules out this
branch through strong coupling and an angular Laplacian instability in the
even-parity sector \cite{Tsujikawa:2022lww}.

We therefore do not rely on degenerate or nonanalytic $X_s\ne0$ branches in the
no-go analysis below.  Instead, we focus on the regular branch connected to
asymptotic flatness, for which
\begin{equation}
X(r_s)=0 .
\label{Xzero}
\end{equation}
\section{Analytic branches and the Schwarzschild solution}
\label{sec:analytic_nohair}

We apply the no-hair reduction to the analytic regular branch. For an asymptotically flat regular BH, the relevant $X=0$ points are
\ba (\phi,X) &\to& (\phi_0,0)\quad (r\to\infty),\qquad (\phi_s,0)\quad (r=r_s),\nonumber \\ & & (\phi_c,0)\quad (r=0), \label{threeXzeroPoints} 
\ea 
where $\phi_0$, $\phi_s$, and $\phi_c$ denote the scalar values at infinity, at the horizon, and at the center, respectively. The endpoints at $r=0$ and $r=\infty$ are not by themselves BH conditions, as they also occur for regular horizonless configurations. The additional BH input is the regular horizon endpoint $X(r_s)=0$, which selects the branch remaining after the generic analytic near-horizon branch with $X_s\ne0$ is obstructed in Sec.~\ref{sec:horizonbranch}.

At infinity we set $\varphi\equiv\phi-\phi_0$.  Asymptotic flatness requires
\begin{equation}
 f\to1,
 \quad h\to1,
 \quad \varphi\to0,
 \quad X\to0
 \qquad (r\to\infty),
\label{asymflatvarphi}
\end{equation}
and the constant-scalar Minkowski vacuum satisfies
$G_2(\phi_0,0)=G_{2,\phi}(\phi_0,0)=0$. 
We assume that all $G_i$ are analytic and regular 
at the three $X=0$ points.  At infinity this means
\begin{equation}
G_i(\phi,X)=\sum_{m,n=0}^{\infty}
g_{i,mn}\,\varphi^m X^n ,
\label{analyticGi}
\end{equation}
where $g_{i,mn}$ are Taylor coefficients and $m,n$ are non-negative integers;
near $r=r_s$ and $r=0$ one replaces $\varphi$ by $\phi-\phi_s$ and
$\phi-\phi_c$, respectively.  Fractional, negative, or logarithmic powers of
$X$ are therefore excluded in this section.

For the metric ansatz \eqref{metric}, the independent background equations 
can be written as
\begin{equation}
{\mathfrak E}_A=0,
\qquad A=t,r,\phi,
\label{bgEqsFormal}
\end{equation}
where $A=t,r$ denote the metric equations and $A=\phi$ the scalar equation.
Each equation is a finite sum of structures
\begin{equation}
{\mathfrak B}^{abI}_{Ai}[f,h,\phi]\,
\partial_\phi^a\partial_X^bG_i(\phi,X) ,
\label{operatorForm}
\end{equation}
where $a$ and $b$ are nonnegative integers, and $I$ labels the finite set of tensorial 
structures. Substituting \eqref{analyticGi} gives
\be
{\mathfrak B}^{abI}_{Ai}
\partial_\phi^a\partial_X^bG_i
=\sum_{m\geq a,\,n\geq b}
(m)_a(n)_b g_{i,mn}{\mathfrak B}^{abI}_{Ai}
\varphi^{m-a}X^{n-b},
\label{expandedOperatorForm}
\ee
where $(m)_a\equiv m!/(m-a)!$ and $(n)_b\equiv n!/(n-b)!$.
At infinity, if
\be
\varphi=O(r^{-\sigma}),\quad
f-1=O(r^{-\mu}),\quad
h-1=O(r^{-\nu})\,,
\label{falloffsGeneral}
\ee
with $\sigma,\mu,\nu>0$, and
${\mathfrak B}^{abI}_{Ai}=O(r^{-w^{abI}_{Ai}})$, then
$X=O(r^{-\chi})$ with $\chi=2\sigma+2$.  
A nonzero term generated by $g_{i,mn}\varphi^mX^n$ scales asymptotically as
$r^{-W_0}$, where
$W_0=w^{abI}_{Ai}+(m-a)\sigma+(n-b)\chi$.
Since both $\varphi$ and $X$ decay at spatial infinity, terms with larger
$m$ or $n$ fall off faster.  At any fixed order in $1/r$, only finitely many
Taylor terms contribute; hence the leading scalar equation is determined by
the finite Taylor jet
\begin{equation}
G_i^{\rm lead}(\phi,X)=
\sum_{(m,n)\in S_i}g_{i,mn}\,\varphi^mX^n,
\label{leadGi}
\end{equation}
so that
\begin{equation}
{\mathfrak E}_\phi^{\rm lead}[G_i]
=
{\mathfrak E}_\phi^{\rm lead}[G_i^{\rm lead}] .
\label{leadIdentity}
\end{equation}
Here $S_i$ is the finite set of index pairs labeling the individual terms in the
Taylor expansion that enter the first nonvanishing scalar equation at the
endpoint under consideration; it can contain more than one pair.

The same finite-jet reduction applies locally at the other endpoints where
$X=0$.  Near the regular center, we have
\be
f=f_c+\cO(r^2),\quad h=1+\cO(r^2),\quad
\phi=\phi_c+\cO(r^2)\,,
\label{centerExpansion}
\ee
so that $\phi'=\cO(r)$ and $X=\cO(r^2)$.
The scalar equation can then be expanded in powers of $r$.  At any fixed order
in this expansion, only finitely many terms contribute.  Hence, near the center,
the relevant part of $G_i$ is a finite Taylor jet of the form \eqref{leadGi},
with $\varphi$ replaced by $\phi-\phi_c$.

Near the regular horizon, the expansion is performed in powers of $r-r_s$,
with $\varphi\to\phi-\phi_s$ and $X(r_s)=0$.  Thus the leading scalar equation
near any $X=0$ endpoint is controlled by finitely many Taylor coefficients.
The center and infinity endpoints are not specific to BHs; the additional BH
input is the regular horizon endpoint $X(r_s)=0$, which separates the problem
into an exterior patch connecting $r=\infty$ to $r_s$ and an interior patch
connecting $r=0$ to the same horizon endpoint.

The background scalar equation can be written as
\begin{equation}
\frac{1}{r^2}\sqrt{\frac{h}{f}}\frac{\rd}{\rd r}
\left(r^2\sqrt{\frac{f}{h}}J^r\right)+P_\phi=0,
\label{ScalarCurrentEq}
\end{equation}
where $J^r$ is the radial current, whose explicit expression is given in
Appendix~\ref{app:logchain}, Eq.~\eqref{JrExplicit}, and $P_\phi$ is defined
in \cite{Minamitsuji:2022vbi}.  
Since the finite jets contain only non-negative powers of $X$, the current
factorizes in each local $X=0$ neighborhood as
\begin{equation}
J^r=h\phi'\left[{\mathfrak A}_q(r,\theta_a)
+\cO(h\phi')\right],
\qquad q=\infty,s,c,
\label{JrRegularExpansion}
\end{equation}
where $\theta_a$ denotes $f,h,\phi$ and 
finitely many radial derivatives, and
${\mathfrak A}_q$ is the regular leading coefficient of $J^r/(h\phi')$ in the
corresponding local expansion.  The bracket is finite at the corresponding
endpoint.  At infinity ${\mathfrak A}_\infty \to
\eta\equiv G_{2,X}(\phi_0,0)$, and we assume $\eta\ne0$ 
so that the scalar mode is not strongly coupled around the vacuum.

For shift-symmetric theories, $P_\phi=0$, and 
Eq.~\eqref{ScalarCurrentEq} gives ${\cal Y}\equiv r^2\sqrt{f/h}\,J^r=C$
in each connected vacuum patch.  In the exterior, regularity of
$J_\mu J^\mu=(J^r)^2/h$ at the $X(r_s)=0$ horizon requires $J^r\to0$ and hence
$C=0$ \cite{Hui:2012qt}.  In the interior, regularity of $J^r$ at the center,
together with \eqref{centerExpansion}, gives ${\cal Y}\to0$ as $r\to0$, so the
interior constant also vanishes.  Thus $J^r=0$ in both patches.  The local
factorization \eqref{JrRegularExpansion} then selects $\phi'=0$ near the
corresponding endpoints.  
On a nondegenerate analytic branch, the uniqueness theorem for regular ordinary
differential equations ensures that the solution with $\phi'=0$ extends
throughout each connected regular patch.  Moving to a branch with $\phi'\ne0$
at a finite radius would require the bracket in \eqref{JrRegularExpansion} to
vanish or become singular, which lies outside the nondegenerate regime.
Hence
\begin{equation}
\phi'(r)=0
\label{phiPrimeZeroPowerLaw}
\end{equation}
for $r>r_s$ and $0<r<r_s$.  The regular horizon endpoint is the extra BH input:
it fixes the exterior current charge and supplies the common endpoint for the
interior argument.

For non-shift-symmetric theories, $P_\phi$ is a genuine source for the current
equation, ${\rm d}{\cal Y}/{\rm d}r=-r^2\sqrt{f/h}\,P_\phi$.
The finite Taylor jets at $q=\infty,s,c$ organize the local endpoint sources,
but by themselves they do not give a fully nonperturbative exclusion of
source-driven branches.  Indeed, near a nondegenerate $X(r_s)=0$ horizon, a
regular finite-jet source $P_\phi=p_\ell(r-r_s)^\ell+\cdots$ with $\ell\ge0$
gives, after the homogeneous current charge is removed,
${\cal Y}=\cO((r-r_s)^{\ell+1})$, which is compatible with finite $\phi'$ and
$X\to0$.

The perturbative non-shift-symmetric exclusion of
\cite{Minamitsuji:2022vbi} assumes more than local analyticity at the $X=0$
endpoints. The zeroth-order theory contains the nondegenerate kinetic term
$G_2=\eta X$ with $\eta\ne0$, while the positive-power corrections are taken
throughout the branch as
$G_I^{\rm pp}=\alpha \tilde\alpha_I(\phi)X^{p_I}$ with integer $p_I\ge0$,
where $\tilde\alpha_I(\phi)$ is analytic and $|\alpha|\ll1$.
For $r>r_s$, the fields 
are expanded around the Schwarzschild branch as
\ba
f&=& f_{\rm S} [1+\sum_{j\ge1}\alpha^j\hat f_j(r)]^2,\qquad
h=f_{\rm S} [1+\sum_{j\ge1}\alpha^j\hat h_j(r)]^{-2},\nonumber \\
\phi &=&
\sum_{j\ge0}\alpha^j\hat\phi_j(r),
\qquad f_{\rm S}=1-\frac{r_s}{r}.
\label{smallAlphaExpansion}
\ea
Since the solution remains perturbatively close to $\hat\phi_0$ and to the
$X=0$ branch, the leading powers in the local expansions at the endpoints
remain unchanged along the branch, except for accidental zeros that shift the
leading term to higher order.  
Regularity at $r=r_s$ removes logarithmic
integration constants order by order, while asymptotic flatness with
$\phi'(\infty)=0$ removes growing particular
solutions \cite{Minamitsuji:2022vbi}. 
Thus the exterior perturbative solution has $\hat f_j=\hat h_j=0$ and
$\hat\phi_j={\rm const}$, fixing analytic horizon data to Schwarzschild values.  Although \eqref{smallAlphaExpansion} is not a regular center ansatz, the $\alpha=0$ equations are the constant-scalar vacuum Einstein equations, so these horizon data have only the Schwarzschild continuation.
By uniqueness of the nondegenerate local analytic branch, the interior perturbative
continuation also has $\hat f_j=\hat h_j=0$ and $\hat\phi_j={\rm const}$.  Since the
resulting Schwarzschild metric is singular 
at $r=0$, no regular BH exists
on the perturbative positive-power branch 
continuously connected to GR.  
A putative large-coupling hairy solution without a Schwarzschild limit as
$\alpha\to0$ would be a disconnected branch, not constrained by this
perturbative exclusion.

Regular horizonless configurations provide a comparison, not an additional
input to the BH proof.  In the shift-symmetric case, the single connected region
has $h>0$ and extends from the regular center to infinity; center regularity
sets $C=0$, and the same nondegenerate-current argument gives $\phi'(r)=0$.
For non-shift-symmetric theories, by contrast, $P_\phi$ sources the scalar
equation, so there is no conserved current flux, and center and infinity
regularity do not in general exclude source-driven branches.  Thus the horizon
endpoint is essential for the BH current argument, while the non-shift-symmetric
case is excluded only on the perturbative positive-power branch described above.

For \eqref{phiPrimeZeroPowerLaw}, the metric equations reduce to the vacuum
Einstein equations with effective reduced Planck mass squared
$2G_4(\phi_0,0)$.  Birkhoff's theorem yields the Schwarzschild branch
\begin{equation}
f=h=1-\frac{2M}{r},
\qquad
\phi=\phi_0,
\label{SchwarzschildFromPowerLaw}
\end{equation}
where the ADM mass $M$ is fixed by matching 
at the horizon. The Kretschmann
scalar constructed from the Riemann tensor $R_{\mu\nu\rho\sigma}$ is
\begin{equation}
R_{\mu\nu\rho\sigma}R^{\mu\nu\rho\sigma}
=\frac{48M^2}{r^6}\,.
\label{Kretschmann}
\end{equation}
Hence $M\ne0$ violates the regular-center requirement, whereas $M=0$ is flat
spacetime and has no horizon. 
A nonsingular asymptotically flat BH is excluded on the $X(r_s)=0$ branch both for nondegenerate shift-symmetric analytic theories and for perturbative positive-power non-shift-symmetric theories connected to GR, the latter by the small-coupling argument above.

\section{Nonanalytic obstructions and 
the sGB completion}
\label{sec:sGBloophole}

We now ask whether a covariantly regular nonanalytic dependence on $X$ can evade
this analytic-branch exclusion on $X(r_s)=0$.  Fractional powers such as
$|X|^{3/2}$ in \eqref{fractionalG3Example} are too soft: they remain multiplied by the explicit
$h\phi'$ in the current.  The marginal possibilities are square roots
and logarithms, whose $X$ derivatives can leave a finite, $\phi'$-independent
term in $J^r$ as $X\to0$ without making $P_\phi$
singular \cite{Sotiriou:2013qea,Babichev:2017guv}:
\ba
\hspace{-0.7cm}
& &
G_2=\eta X+\alpha_2(\phi)\sqrt{|X|},
\qquad
G_3=\alpha_3(\phi)\ln|X|,\nonumber \\
\hspace{-0.7cm}
& &
G_4=\frac{M_{\rm Pl}^2}{2}+\alpha_4(\phi)\sqrt{|X|},
\qquad
G_5=\alpha_5(\phi)\ln|X|.
\label{marginalExamples}
\ea
More singular nonanalyticities make $J^r$ or $P_\phi$ divergent at $X=0$,
whereas milder ones remain within the regular factorization.  The square-root
terms in $G_2$ and $G_4$ select the generically obstructed $X_s\ne0$ branch
discussed in Sec.~\ref{sec:horizonbranch}.  The $G_3$ logarithm can keep
$X_s=0$, but fails the asymptotically flat power-law test unless its relevant
coefficients vanish order by order \cite{Minamitsuji:2022vbi}.

The only remaining marginal possibility is the quintic logarithm.  In the static
current \eqref{JrExplicit}, the $\phi'^2G_{5,X}$ terms scale, up to regular
metric factors, as $-(2X/h)G_{5,X}$.  A finite $\phi'$-independent contribution
therefore requires $XG_{5,X}$ to approach a finite nonzero value.  This current
criterion is insufficient: covariance also requires all logarithmic and
inverse-$X$ structures in the equations of motion to cancel.  We thus consider
\begin{equation}
G_5=A_5(\phi,X)\ln |X|+g_5(\phi,X),
\label{generalQuinticLog}
\end{equation}
where $A_5$ and $g_5$ are regular near $X=0$.  Appendix~\ref{app:logchain}
shows that only $A_5(\phi,0)$ is marginal in the current; we denote it by
$A_5(\phi)$ below.

A quintic logarithm with nonconstant $A_5(\phi)$ is not covariantly regular by
itself.  The Horndeski equations contain logarithmic and inverse-$X$ terms.
Requiring their cancellation fixes the lower singular functions uniquely, as
in Eq.~\eqref{uniqueLogChain}, up to regular analytic functions and total
derivatives.  Appendix~\ref{app:logchain} derives this chain from
\eqref{generalQuinticLog}, without assuming the sGB form in advance.  Setting
$A_5=-4\xi_{,\phi}$, Eq.~\eqref{uniqueLogChain} is precisely the Horndeski
representation of the covariant sGB 
term 
$\int \rd^4x\sqrt{-g}\,\xi(\phi)\GB$ \cite{Kobayashi:2011nu,Langlois:2022eta},
whose equations of motion are finite:
\begin{equation}
{\cal E}_\phi^{\rm sGB}=\xi_{,\phi}\GB,
\qquad
{\cal E}_{\mu\nu}^{\rm sGB}
=-4P_{\mu\rho\nu\sigma}\nabla^\rho\nabla^\sigma\xi,
\label{sGBFiniteEOM}
\end{equation}
up to convention-dependent signs, where $P_{\mu\rho\nu\sigma}$ is the double dual
of the Riemann tensor.  The singular individual Horndeski functions are only
representation artifacts.  Any marginal nonanalytic term outside the chain
\eqref{uniqueLogChain} leaves uncancelled logarithmic or inverse-$X$ structures
in the covariant equations, or in the quadratic action, and is tested by
asymptotic flatness at $X=0$.  Thus the sGB chain is the unique covariantly
regular marginal nonanalytic completion of the local $X=0$ reduction.

This completion is not a regular-BH counterexample.
Appendix~\ref{app:sGBcenter} shows that, for the Einstein--sGB model and its analytic Horndeski completions with a nondegenerate scalar kinetic sector, the regular-center branch connected to a constant-scalar Minkowski center remains locally Minkowski and cannot develop nonzero mass or hair without encountering a degeneracy or singularity. 
Known asymptotically flat sGB hairy 
BHs possess regular horizons, but their interiors terminate at a curvature singularity 
rather than at a regular center \cite{Sotiriou:2014pfa}.

A related scalar--tensor construction arises from the conformally regularized
infinite Lovelock tower \cite{Colleaux:2020wfv,Fernandes:2025curvature}.
At order $n\geq2$,
\begin{equation}
\begin{aligned}
G_2^{(n)}(X)&=2^{n+1}(n-1)(2n-3)X^n,\\
G_3^{(n)}(X)&=-2^n n(2n-3)X^{n-1},\\
G_4^{(n)}(X)&=2^{n-1}nX^{n-1},\\
G_5^{(n)}(X)&=
\begin{cases}
-4\ln |X|, & n=2,\\
-\dfrac{2^{n-1}n(n-1)}{n-2}X^{n-2}, & n\geq3,
\end{cases}
\end{aligned}
\label{LovelockGi}
\end{equation}
up to the coefficient multiplying each Lovelock density. In the notation of
Appendix~\ref{app:sGBcenter}, $G_{2,X}^{(n)}(0)=0$, while shift symmetry gives
$G_{3,\phi}=q_*=0$, and hence $\eta_*=Z_*=0$. 
The $n\geq3$ sector is analytic at $X=0$ but generates no quadratic kinetic term for scalar perturbations, whereas the $n=2$ contribution adds the nonanalytic linear sGB chain without lifting this 
degeneracy. Thus the tower lies outside the nondegenerate flat-vacuum class of
Appendix~\ref{app:sGBcenter}. Since its quadratic scalar kinetic term vanishes
at the $X=0$ vacuum, it is strongly coupled there and does not provide a
controlled spherical regular-BH loophole within the four-dimensional
Horndeski class considered here. The known planar regular BHs in this tower 
are not covered by the spherical-center argument because $X$ remains finite 
under the corresponding boundary conditions \cite{Fernandes:2025curvature}. 
However, on these solutions the even-parity scalar mode is strongly coupled, 
and the odd-parity sector develops ghost or Laplacian instabilities 
near the origin \cite{DeFelice:2025fzv}.

\section{Conclusions}

We have established no-go results for static, spherically symmetric regular BHs
with a time-independent scalar in four-dimensional Horndeski theories. For
$X(r_s)\ne0$, the horizon generically develops divergent propagation speeds or
ghost/Laplacian instabilities, apart from special simultaneous degeneracies not
covered here. On the regular branch $X(r_s)=0$, analyticity reduces the leading scalar equation at each endpoint to a finite Taylor jet. In nondegenerate shift-symmetric
theories, the conserved current enforces a constant scalar and the Schwarzschild
metric; nonzero ADM mass therefore implies a central singularity. For
non-shift-symmetric positive-power couplings, the same exclusion holds on the
small-coupling branch connected to Schwarzschild. Thus spherical regular BHs
are absent in these analytic sectors, rather than being destabilized only near
the center as in NED \cite{DeFelice:2024seu}.

Marginal nonanalytic terms also fail to provide a controlled escape. On the
$X(r_s)=0$ branch, asymptotic flatness and covariant regularity single out the
quintic logarithm and fix its completion uniquely to the sGB chain derived in
Appendix~\ref{app:logchain}. Other isolated logarithmic, fractional, or
inverse-power terms are excluded by the horizon analysis, asymptotic flatness,
or covariant regularity.
Appendix~\ref{app:sGBcenter} shows that the nondegenerate sGB branch connected
to a smooth flat center is locally Minkowski with a constant scalar, while
known hairy BH interiors terminate at curvature singularities. The related
Lovelock tower lies outside the nondegenerate class and is strongly coupled at
$X=0$.

Our results provide a sharp benchmark for regular-BH constructions with static scalar 
profiles in Horndeski theories and identify the assumptions that must be relaxed 
in viable extensions. Time-dependent 
shift-symmetric branches,
$\phi=qt+\psi(r)$ \cite{Babichev:2013cya,Kobayashi:2014eva}, regular BHs in
beyond-Horndeski/DHOST theories
\cite{Chagoya:2018lmv,Babichev:2020qpr,Baake:2021qja}, and geometrically regular
BHs with hedgehog scalar hair \cite{Bahamonde:2026bvh} lie outside these
assumptions. Whether these broader constructions can yield fully stable regular
BHs with a healthy perturbative sector remains open.

\section*{Acknowledgements}

S.T. acknowledges support from JSPS KAKENHI Grant Nos.~26K07090 and 26H00847, and from the Waseda University Special Research 
Projects (No.~2026C-486).

\appendix

\section{Derivation of the logarithmic chain from regularity}
\label{app:logchain}

We derive the logarithmic chain, whose final form is Eq.~\eqref{uniqueLogChain},
directly from the marginal quintic logarithm. We write
\begin{equation}
G_5=A_5(\phi,X)L+g_5(\phi,X),
\qquad L\equiv\ln |X|,
\label{appGeneralG5}
\end{equation}
and use ``regular'' to mean Taylor expandable in $X$ around $X=0$. To identify
the part of \eqref{appGeneralG5} that can evade the analytic current
factorization, we recall the radial current used in the main text:
\ba
\hspace{-0.2cm}
J^r &=& h \phi' \bigg[ G_{2,X}
-\frac{4f+rf'}{2rf}h \phi' G_{3,X} \nonumber \\
\hspace{-0.2cm}
& &
+2\frac{f(1-h)-hrf'}{r^2 f} G_{4,X}
+2h \phi'^2 \frac{fh+rf'h}{r^2 f} G_{4,XX} \nonumber \\
\hspace{-0.2cm}
& &-\frac{f'h (1-3h)\phi'}{2r^2 f}G_{5,X}
-\frac{f' h^3 \phi'^3}{2r^2 f}G_{5,XX} \bigg].
\label{JrExplicit}
\ea
The regular part $g_5$ belongs to the analytic sector and only modifies the regular coefficient multiplying $h \phi'$ in
\eqref{JrExplicit}. It therefore cannot generate a finite
$\phi'$-independent contribution 
on the $X \to 0$ branch.

We first show which part of $A_5(\phi,X)L$ is marginal.  
Since $A_5$ is regular in $X$, it can be decomposed as
\begin{equation}
A_5(\phi,X)=A_0(\phi)+X\bar A_5(\phi,X),
\qquad A_0(\phi)\equiv A_5(\phi,0),
\label{appA5split}
\end{equation}
with regular $\bar A_5$.  Since $\partial_XL=1/X$, the first term gives
$X\partial_X(A_0L)=A_0$ and
$X^2\partial_X^2(A_0L)=-A_0$, and hence can leave finite contributions
to the current. The second term is subleading. 
Indeed,
\begin{align}
X\partial_X(X\bar A_5L)&={\cal O}(XL),
\nonumber\\
X^2\partial_X^2(X\bar A_5L)&={\cal O}(XL)+{\cal O}(X),
\label{appSubleadingXlog}
\end{align}
which vanish in the $X\to0$ current. 
Thus $X\bar A_5L$ cannot provide an independent escape from 
the regular no-hair branch.  The only marginal quintic logarithm 
is therefore
\begin{equation}
G_5^{\rm marg}=A_0(\phi)L.
\label{appMarginalG5}
\end{equation}
Below we rename $A_0(\phi)$ as $A_5(\phi)$.

It remains to determine whether this marginal term can be embedded 
in covariantly regular Horndeski theories.  We use the trace notation
\begin{align}
[\phi]&\equiv\Box\phi,
\qquad
{\cal P}_2\equiv[\phi]^2-[\phi^2],
\nonumber\\
{\cal P}_3&\equiv[\phi]^3-3[\phi][\phi^2]+2[\phi^3].
\label{appTraceNotation}
\end{align}
The relevant Horndeski terms are
\begin{equation}
\begin{aligned}
{\cal L}_3&=-G_3[\phi],\\
{\cal L}_4&=G_4R+G_{4,X}{\cal P}_2,\\
{\cal L}_5&=G_5G_{\mu\nu}\phi^{\mu\nu}
-\frac{G_{5,X}}{6}{\cal P}_3 .
\end{aligned}
\label{appHorndeskiPieces}
\end{equation}
The lower functions must remove the independent nonregular structures produced
by $A_5(\phi)L$.  Power counting fixes their possible $X$ dependence.  A term
$X^nL$ in $G_4$ contributes $X^nLR$ and, through $G_{4,X}$,
$X^{n-1}L{\cal P}_2$; only $n=1$ can eliminate both the $XLR$ and
$L{\cal P}_2$ structures generated by the quintic term.  Similarly, only
$X L$ in $G_3$ can absorb the $XL[\phi]$ structure, and only $X^2L$ in $G_2$
can remove the $X^2L$ structure.  All other logarithmic powers either generate
new uncancelled nonregular terms or are subleading in the marginal 
cancellation. Hence the most general lower completion relevant at this order is
\begin{equation}
\begin{aligned}
G_5^{\rm sing}&=A_5(\phi)L,\\
G_4^{\rm sing}&=X\left[B_4(\phi)L+C_4(\phi)\right],\\
G_3^{\rm sing}&=X\left[B_3(\phi)L+C_3(\phi)\right],\\
G_2^{\rm sing}&=X^2\left[B_2(\phi)L+C_2(\phi)\right].
\end{aligned}
\label{generalLogAnsatz}
\end{equation}
The powers of $X$ in this ansatz are therefore fixed by regularity, not by
assuming the sGB form. 

Substituting \eqref{generalLogAnsatz} into the full Horndeski integrand in
the action (1), with the relevant ${\cal L}_3$--${\cal L}_5$ pieces displayed
in \eqref{appHorndeskiPieces}, gives the nonanalytic contribution generated by
$G_i^{\rm sing}$ $(i=2,\ldots,5)$:
\begin{align}
{\cal L}_{\rm sing}={}&
A_5LG_{\mu\nu}\phi^{\mu\nu}
-\frac{A_5}{6X}{\cal P}_3
\nonumber\\
&+X(B_4L+C_4)R
+\bigl[B_4(L+1)+C_4\bigr]{\cal P}_2
\nonumber\\
&-X(B_3L+C_3)[\phi]
+X^2(B_2L+C_2).
\label{LsingGeneral}
\end{align}
The nonregular content of the first line can be made explicit by integrating by
parts and using $\nabla_\mu G^{\mu\nu}=0$, $\nabla_\mu X=-\phi_{\mu\nu}\nabla^\nu\phi$,
and the commutator of covariant derivatives.  Here and below, $\doteq$
denotes equality up to a total derivative in the action.  We then obtain
\begin{align}
&A_5LG_{\mu\nu}\phi^{\mu\nu}
-\frac{A_5}{6X}{\cal P}_3
\nonumber\\
&\doteq {\cal R}[A_5]
+A_{5,\phi}X(2-L)R
+A_{5,\phi}(1-L){\cal P}_2
\nonumber\\
&\quad
-A_{5,\phi\phi}X(7-3L)[\phi]
+2A_{5,\phi\phi\phi}X^2(3-L),
\label{quinticIBPRegularRemainder}
\end{align}
where ${\cal R}[A_5]$ is regular as $X\to0$.  This identity displays all
logarithmic and inverse-$X$ structures generated by the marginal quintic
logarithm.

Combining Eqs.~\eqref{LsingGeneral} and
\eqref{quinticIBPRegularRemainder}, the nonregular remainder is
\begin{align}
\Delta {\cal L}_{\rm nreg}={}&
X\left[(B_4-A_{5,\phi})L+C_4+2A_{5,\phi}\right]R
\nonumber\\
&+\left[(B_4-A_{5,\phi})L+B_4+C_4+A_{5,\phi}\right]{\cal P}_2
\nonumber\\
&+X\left[(-B_3+3A_{5,\phi\phi})L
-(C_3+7A_{5,\phi\phi})\right][\phi]
\nonumber\\
&+X^2\left[(B_2-2A_{5,\phi\phi\phi})L
+C_2+6A_{5,\phi\phi\phi}\right].
\label{DeltaLnreg}
\end{align}
For arbitrary smooth configurations, the structures $R$, ${\cal P}_2$,
$[\phi]$, and the scalar term are kinematically independent, since they
represent distinct covariant structures.  Regular analytic functions cannot
cancel nonanalytic logarithms or inverse powers of $X$ for generic field
configurations.  Covariant regularity therefore requires each coefficient in
\eqref{DeltaLnreg} to vanish separately:
\begin{equation}
\begin{aligned}
B_4&=A_{5,\phi},
& C_4&=-2A_{5,\phi},\\
B_3&=3A_{5,\phi\phi},
& C_3&=-7A_{5,\phi\phi},\\
B_2&=2A_{5,\phi\phi\phi},
& C_2&=-6A_{5,\phi\phi\phi}.
\end{aligned}
\label{logCoeffConditions}
\end{equation}
Substituting these coefficients into \eqref{generalLogAnsatz}, we obtain
\begin{equation}
\begin{aligned}
G_5^{\rm sing} &= A_5(\phi)\ln |X|,\\
G_4^{\rm sing} &=-A_{5,\phi}(\phi)X\left(2-\ln |X|\right),\\
G_3^{\rm sing} &=-A_{5,\phi\phi}(\phi)X\left(7-3\ln |X|\right),\\
G_2^{\rm sing} &=-2A_{5,\phi\phi\phi}(\phi)X^2
\left(3-\ln |X|\right).
\end{aligned}
\label{uniqueLogChain}
\end{equation}
The chain is unique up to
regular analytic functions and total derivatives.  Any change in the relative
coefficients, or any isolated marginal nonanalytic term outside the chain,
leaves at least one independent logarithmic or inverse-$X$ structure in
\eqref{DeltaLnreg}.

Finally, the regular remainder can be identified only after the chain has been
fixed. With $A_5=-4\xi_{,\phi}$, Eq.~\eqref{uniqueLogChain} is equivalent, up
to a total derivative and convention-dependent signs, to the sGB interaction
\begin{equation}
\int \rd^4x\sqrt{-g}\,\xi(\phi)R_{\rm GB}^2\,,
\end{equation}
where
\be
R_{\rm GB}^2=R^2-4R_{\mu\nu}R^{\mu\nu}
+R_{\mu\nu\rho\sigma}R^{\mu\nu\rho\sigma}\,.
\ee
Here $R_{\mu\nu}$ is the Ricci tensor.
Thus regularity first fixes the logarithmic chain; only afterward can its
covariant completion be identified with the sGB combination.

\section{Vacuum regular center in sGB and analytic completions}
\label{app:sGBcenter}

This appendix presents the local regular-center argument used in
Sec.~\ref{sec:sGBloophole}.  We first consider the minimal Einstein--sGB
model
\begin{equation}
{\cal S}_{\rm EsGB}=\int \rd^4x\sqrt{-g}\left[
\frac{M_{\rm Pl}^2}{2}R+\eta X+\xi(\phi)\GB
\right],
\label{appESGBaction}
\end{equation}
without a scalar potential, cosmological constant, or matter, and then extend
the result to nondegenerate Horndeski completions that admit a constant-scalar
Minkowski vacuum and whose functions are analytic at
$(\phi,X)=(\phi_c,0)$.  We assume $\eta\ne0$, with canonical normalization
corresponding to $\eta=1$.  The same regular-center expansion is used for
neutron stars in Ref.~\cite{Minamitsuji:2022tze}; the vacuum equations below
correspond to the regular branch obtained by taking the central density to
zero.

For the metric ansatz \eqref{metric}, a smooth spherical center requires
\begin{equation}
\begin{aligned}
f(r)&=f_c\left(1+f_2r^2+\cdots\right),\\
h(r)&=1+h_2r^2+\cdots,\\
\phi(r)&=\phi_c+\phi_2r^2+\cdots .
\end{aligned}
\label{appESGBcenterexp}
\end{equation}
Thus $\phi'(r)=\cO(r)$ and $X=\cO(r^2)\to0$.  The constant term
$\xi(\phi_c)\GB$ is topological in four dimensions, while derivatives of
$\xi$ enter the metric equations through $\nabla_\mu\nabla_\nu\xi$ and the
scalar equation through $\xi_{,\phi}\GB$.  Substituting
\eqref{appESGBcenterexp} into the two independent metric equations and the
scalar equation gives, at leading order,
\begin{align}
0&=h_2\left(3M_{\rm Pl}^2-48\xi_1\phi_2\right),
\label{appESGBeq1}\\
0&=M_{\rm Pl}^2\left(2f_2+h_2\right)-32\xi_1f_2\phi_2,
\label{appESGBeq2}\\
0&=6\eta\phi_2+24\xi_1f_2h_2,
\label{appESGBeq3}
\end{align}
where $\xi_1=\xi_{,\phi}(\phi_c)$; the signs of the $\xi_1$ terms depend on
conventions.  Equations \eqref{appESGBeq1}--\eqref{appESGBeq3} force
\begin{equation}
f_2=h_2=\phi_2=0.
\label{appESGBsecondzero}
\end{equation}
Indeed, if $h_2=0$, then \eqref{appESGBeq3} gives $\phi_2=0$, since
$\eta\ne0$, and \eqref{appESGBeq2} gives $f_2=0$.  If $h_2\ne0$, then
\eqref{appESGBeq1} fixes $16\xi_1\phi_2=M_{\rm Pl}^2$ for $\xi_1\ne0$,
and \eqref{appESGBeq2} reduces to $M_{\rm Pl}^2h_2=0$, a contradiction;
for $\xi_1=0$, Eq.~\eqref{appESGBeq1} already gives $h_2=0$.

The argument then iterates.  Suppose all lower coefficients vanish and that
$r^N$, with $N>2$, is the first possible nonzero order,
\begin{equation}
\begin{aligned}
f(r)&=f_c\left(1+f_Nr^N+\cdots\right),\\
h(r)&=1+h_Nr^N+\cdots,\\
\phi(r)&=\phi_c+\phi_Nr^N+\cdots .
\end{aligned}
\label{appESGBNexp}
\end{equation}
Curvature and second scalar derivatives scale as $r^{N-2}$, so the sGB
contributions begin at $r^{2N-4}$ and are subleading at order $r^{N-2}$.
The leading system therefore reduces to its Einstein--scalar form,
\begin{equation}
\begin{aligned}
0&=M_{\rm Pl}^2(N+1)h_N,\\
0&=M_{\rm Pl}^2\left(Nf_N+h_N\right),\\
0&=\eta N(N+1)\phi_N .
\end{aligned}
\label{appESGBrecursion}
\end{equation}
These equations again give $f_N=h_N=\phi_N=0$.  Hence the analytic vacuum
regular-center branch is locally
\begin{equation}
f=f_c,
\qquad
h=1,
\qquad
\phi=\phi_c,
\label{appESGBlocallyflat}
\end{equation}
which is Minkowski space after a constant rescaling of time.

We now extend the result.  Let $G_i^{\rm reg}$ denote the functions defining
the total regular analytic Horndeski sector, including the terms
$M_{\rm Pl}^2R/2+\eta X$.  We assume that this sector admits the same
constant-scalar flat vacuum,
\begin{equation}
\begin{aligned}
G_2^{\rm reg}(\phi_c,0)&=G_{2,\phi}^{\rm reg}(\phi_c,0)=0,\\
M_*^2&\equiv 2G_4^{\rm reg}(\phi_c,0)>0,
\end{aligned}
\label{appRegHorVacuum}
\end{equation}
and has a nonzero scalar kinetic coefficient after diagonalizing the linear
scalar--metric mixing.  At the order relevant to the leading center equations,
the expansion about $(\phi,X)=(\phi_c,0)$ takes the form
\begin{equation}
{\cal L}_{\rm reg}^{\rm lead}
=\frac{M_*^2}{2}R+q_*\delta\phi R+\eta_*X+\cdots,
\label{appRegHorLinearLag}
\end{equation}
where $\delta\phi\equiv\phi-\phi_c$,
$q_*\equiv G_{4,\phi}^{\rm reg}(\phi_c,0)$, and $\eta_*$ is the effective
regular kinetic coefficient, including, for example, the
$G_{2,X}^{\rm reg}$ and $G_{3,\phi}^{\rm reg}$ contributions in the present
convention.  After diagonalizing the $q_*\delta\phi R$ mixing, the scalar
kinetic coefficient is
\begin{equation}
Z_*\equiv \eta_*+\frac{6q_*^2}{M_*^2}\ne0 .
\label{appRegHorNondeg}
\end{equation}
If $Z_*=0$, the quadratic scalar kinetic term, and hence the linear principal
part, vanishes at the flat center, so higher-order interactions control the
fluctuations.  This is a degenerate, generically strongly coupled scalar sector
outside the present perturbative analysis.

If $r^N$ is the first possible nonzero order in the deviations from the flat
center, the equations linear in $(f_N,h_N,\phi_N)$ are
\begin{align}
0&=M_*^2 G_{\mu\nu}^{(N)}
+2q_*\left(\eta_{\mu\nu}\Box-\partial_\mu\partial_\nu\right)
\delta\phi_N,
\label{appRegHorLinMetric}\\
0&=\eta_*\Box\delta\phi_N+q_*R^{(N)},
\label{appRegHorLinScalar}
\end{align}
where $\delta\phi_N=\phi_Nr^N$, $G_{\mu\nu}^{(N)}$ and $R^{(N)}$ are the
linearized curvatures generated by $(f_N,h_N)$, and derivatives are evaluated
in local Cartesian coordinates at the flat center.  Taking the trace of
\eqref{appRegHorLinMetric} and using \eqref{appRegHorLinScalar} gives
\begin{equation}
Z_*\Box\delta\phi_N=0.
\label{appRegHorDiagScalar}
\end{equation}
Since $\Box r^N=N(N+1)r^{N-2}$ for a static radial function in locally flat
space, $Z_*\ne0$ implies $\phi_N=0$.  The metric equations then reduce to the
linearized vacuum Einstein equations, giving $h_N=f_N=0$.

Relative to these linear $r^{N-2}$ terms, the omitted analytic terms are
either lower-derivative terms, which are subleading by at least two powers of
$r$, or nonlinear derivative and curvature terms, which are at least quadratic
in the first nonzero coefficients and begin no earlier than $r^{2N-4}$.
Both classes are subleading for $N>2$.  For $N=2$, the nonlinear terms can
enter at the same radial order but are at least quadratic in
$\boldsymbol{c}_2\equiv(f_2,h_2,\phi_2)$.  They therefore leave the nonsingular
linear coefficient matrix unchanged, so $\boldsymbol{c}_2=0$ is the unique
nearby solution for $M_*^2>0$ and $Z_*\ne0$.
Hence no regular analytic Horndeski completion satisfying
\eqref{appRegHorVacuum} and \eqref{appRegHorNondeg} can deform the local sGB
branch into a center with nonzero curvature or scalar hair.  If instead
$G_2^{\rm reg}(\phi_c,0)\ne0$,
$G_{2,\phi}^{\rm reg}(\phi_c,0)\ne0$, or $Z_*=0$, this argument does not
apply: these cases correspond, respectively, to a nonzero local cosmological
constant, a scalar tadpole, or a degenerate, generally strongly coupled scalar
sector rather than the nondegenerate flat-vacuum class considered here.

Once written in variables regular at the center, the nondegenerate radial equations have a unique local continuation, so the Minkowski--constant-scalar solution cannot connect to a branch with nonzero hair or mass without encountering a degeneracy or singularity. Therefore, a nonzero-mass static spherical BH in the minimal Einstein--sGB model or such an analytic Horndeski completion cannot have a smooth analytic center.

\bibliographystyle{mybibstyle}
\bibliography{bib}

\end{document}